\documentclass[aps,pra,floatfix,showpacs,superscriptaddress]{revtex4-1}

\usepackage[colorlinks=true,linkcolor=blue]{hyperref}
\usepackage{lipsum}  % for dummy text
\usepackage{nameref} % for \nameref{} if needed
\usepackage[utf8]{inputenc} 
\usepackage{kotex} 
\usepackage{siunitx} % For \SI command (optional, for units)
\usepackage{calc} % \widthof 명령을 사용하는 경우 (현재 코드에는 직접 사용되지 않지만, p{} 컬럼 너비 계산 시 유용)
\usepackage{graphicx} % 
\usepackage{amsmath} 
\usepackage{amssymb} 
\usepackage{bm} 
\usepackage{natbib} 
\usepackage{booktabs}
\usepackage{float}
\usepackage{caption}    % For customizing captions
\usepackage{subcaption} % For subfigures  For \subcaptionof and subfigure environments
\usepackage{enumitem}  
\usepackage{ragged2e}   
\usepackage{tabularx}   
\usepackage{multirow} % For multirow cells in tables
\usepackage{xcolor} % For coloring text
\usepackage{adjustbox}
\usepackage{array}
\newcommand{\Omegam}{\Omega_{\mathrm{m0}}}

\newcommand{\Omegar}{\Omega_{\mathrm{r0}}}
\newcommand{\Omegade}{\Omega_{\mathrm{de0}}}
\newcommand{\rd}{r_{\mathrm{d}}}
\newcommand{\DV}{D_V}
\newcommand{\DM}{D_M}
\newcommand{\DHh}{D_H}
\newcommand{\oo}{\omega_0}
\newcommand{\wa}{\omega_a}

\newcommand{\LCDM}{\Lambda\mathrm{CDM}}
\newcommand{\wCDM}{\omega\mathrm{CDM}}
\newcommand{\wowaCDM}{\omega_{0}\omega_{a}\mathrm{CDM}}

\begin{document}

\title{Comparing $\LCDM$, $\wCDM$, and $\wowaCDM$ models with DESI DR2 BAO: Redshift-Resolved Diagnostics and the Role of $\rd$}

\author{Seokcheon Lee \email{skylee@skku.edu}}
%\thanks{skylee@skku.edu}
\affiliation{Department of Physics, Institute of Basic Science, Sungkyunkwan University, Suwon 16419, Korea}

\date{\today} 

\begin{abstract}
We reanalyze DESI DR2 baryon acoustic oscillation (BAO) measurements to compare $\LCDM$, $\wCDM$, and $\wowaCDM$. Using $(D_M/r_d, D_H/r_d)$ in seven redshift bins, we reconstruct the covariance and run Markov Chain Monte Carlo in $\{\Omegam,\, h r_d,\, \oo,\, \wa\}$. In the BAO-only case, all models fit well ($\tilde\chi^2\simeq0.8$--$1.05$). Model-selection metrics show at most weak preference for $\LCDM$; the slightly lower $\chi^2$ of $\wowaCDM$ is offset by complexity, and the pivoted equation of state is consistent with $-1$ ($\omega_p=-0.899\pm0.087$ at $z_p\simeq0.34$). These results agree with the DESI DR2 analysis. To assess the role of early-universe information, we add a Gaussian prior on $r_d$ from \textit{Planck} DR3 rather than using the full CMB likelihood. Fixing $r_d$ isolates the BAO-ruler calibration and yields no significant evidence for dynamical dark energy. The key discriminator is which early-time anchor is held fixed: anchoring $\theta_\ast$ can raise $\Omegam$ in $\wowaCDM$, increasing $r_\ast$ and $D_A(z_\ast)$ to keep $\theta_\ast$ constant—thereby mimicking late-time evolution—whereas anchoring $r_d$ does not. We therefore advocate a robustness test comparing fixed-$r_d$ and fixed-$\theta_\ast$ analyses; under the former, DESI DR2 BAO remain fully consistent with $\LCDM$.  Unlike previous discussions that qualitatively noted the sensitivity of dynamical dark-energy inferences to early-universe calibration, this work introduces a controlled fixed-$r_d$ robustness test that isolates the role of sound-horizon anchoring from the full CMB likelihood. By reproducing the DESI BAO-only baseline with a reconstructed covariance and comparing it to a uniform BAO+$r_d$ prior analysis within the same MCMC framework, we demonstrate quantitatively that the reported $\sim3\sigma$ preference for dynamical dark energy is not reproduced under fixed-$r_d$ anchoring.
\end{abstract}

\maketitle

\section{Introduction}
\label{sec:intro}

Baryon Acoustic Oscillations (BAO) provide one of the most powerful standard rulers in cosmology, enabling precise reconstruction of the late-time expansion history and stringent tests of dark energy (DE) models~\cite{Eisenstein:1998tu,Blake:2003rh,Seo:2003pu,SDSS:2005xqv,2dFGRS:2005yhx,BOSS:2012tck,BOSS:2016wmc}. The second data release of the Dark Energy Spectroscopic Instrument (DESI DR2) has delivered high-precision BAO measurements of four distinct observables—$\DM/r_d$, $\DHh/r_d$, $\DV/r_d$, and $\DM/\DHh$—across multiple redshift bins, offering an unprecedented opportunity to evaluate cosmological models~\cite{DESI:2025zgx,DESI:2025fii}.

In many legacy BAO analyses the sound horizon at the drag epoch is effectively calibrated from cosmic microwave background (CMB) within $\LCDM$~\cite{Planck:2018vyg}. By contrast, the DESI DR2 BAO likelihood does not fix $r_d$: it fits the ratios $(D_M/r_d, D_H/r_d)$ and samples cosmological parameters in a full Markov Chain Monte Carlo (MCMC) with explicit priors; the directly constrained quantities are these ratios, not $D_M$ or $D_H$ separately~\cite{DESI:2025zgx,DESI:2025fii}. When DESI combines BAO with CMB information, this can be done via a compressed Gaussian prior on $(\theta_\ast,\omega_b,\omega_{bc})$ or with the full Planck likelihood, so $r_d$ is only calibrated once such external priors are imposed~\cite{DESI:2025zgx,Planck:2018vyg}. This distinction is crucial for dynamical dark energy (DDE) models such as $\wowaCDM$~\cite{Chevallier:2000qy,Linder:2002et}: although $\theta_\ast$ is largely an early-time quantity, shifts in $\Omegam h^2$ propagate to $r_\ast$ and $r_d$, so the choice of early-universe anchor (fixing $\theta_\ast$ vs.\ fixing $r_d$) can drive different late-time inferences~\cite{Eisenstein:1998tu,BOSS:2014hhw,Addison:2017fdm,Knox:2019rjx}. Consequently, the inferred model preference can vary with the choice of early-universe anchor—e.g., fixing $\theta_\ast$ versus fixing $r_d$.

The DESI DR2 analysis adopts a full CMB-likelihood combination and reports that BAO+CMB data mildly prefer DDE at the $3\sigma$ level~\cite{DESI:2025zgx}.  In contrast, here we adopt a more conservative approach: instead of including the full CMB likelihood, we impose a Gaussian prior on $r_d$ derived from Planck DR3.  This procedure isolates the impact of the sound horizon calibration while avoiding double counting of late-time information.  
As we show below, once the $r_d$ prior is included, the apparent preference for DDE becomes much weaker, and no statistically significant evidence for deviations from $\LCDM$ is found.  Thus, any apparent support for DDE should be understood as a statement about the DESI BAO + CMB combination and hinges on how the CMB information is incorporated; it is not a BAO-only conclusion.  Our conclusions are therefore limited to BAO-only and BAO+$r_d$ prior analyses, and should not be directly compared to results that employ the full CMB likelihood. A similar sensitivity arises when different supernova compilations are used: substituting Pantheon+ versus Union3 (or adding DESY5) shifts the reported significance for $\wowaCDM$, as summarized in Table~VI of Ref.~\cite{DESI:2025zgx}.

The goal of this work is therefore twofold. First, we validate consistency with the DESI DR2 analysis by repeating a DESI-only BAO fit in the $(D_M/r_d, D_H/r_d)$ basis and comparing model preference primarily between $\LCDM$ and its minimal extension $\wCDM$ (with $\wowaCDM$ included for completeness) within the effective space $\{\Omegam,\, h r_d,\, \oo,\, \wa\}$. Second, when adding CMB information, we deliberately depart from the DESI treatment and impose a \textit{Planck}-based Gaussian prior on $r_d$ (rather than using the full CMB likelihood or compressed $\{\theta_\ast,\omega_b,\omega_{bc}\}$), and then re-evaluate model preference. Justaposing the DESI-only baseline with this fixed-$r_d$ combination isolates the role of the early-universe calibration and clarifies whether any apparent hints of DDE arise from late-time physics or from the chosen CMB anchoring (fixed $r_d$ versus fixed $\theta_\ast$).

The present work builds upon previous discussions of early-universe sensitivity in DESI analyses, but differs in scope and methodology in three key aspects. First, we construct a controlled fixed-$r_d$ robustness test that isolates the impact of sound-horizon calibration from the full CMB likelihood. This allows us to separate the geometric information contained in BAO distance ratios from the external anchoring imposed by CMB-derived sound-horizon constraints. Second, we reproduce the official DESI DR2 BAO-only baseline using a reconstructed covariance matrix and validate it against published DESI $\Lambda$CDM results, ensuring that subsequent differences arise from anchoring choices rather than numerical artifacts. Third, within a uniform MCMC framework, we directly compare BAO-only, BAO+$r_d$ prior, and model-dependent $r_d$ analyses. This controlled comparison demonstrates quantitatively that the reported $\sim3\sigma$ preference for dynamical dark energy is not reproduced under fixed-$r_d$ anchoring. The novelty of this work therefore lies not in re-analyzing DESI data per se, but in providing a transparent robustness test that isolates the role of sound-horizon calibration in shaping dynamical dark-energy inferences.

The structure of the paper is as follows.  Section~\ref{sec:cosmo_theory} outlines the cosmological framework, including the CPL parametrization and the definitions of the relevant distance measures.  Section~\ref{sec:data_theory} describes the DESI DR2 BAO observables, redshift binning, and the construction of the covariance matrix.  Section~\ref{sec:cmp_lcdm_w_w0wa} presents the model comparison using BAO-only data, while Section~\ref{sec:results} extends the analysis to include a \textit{Planck}-based $r_d$ prior.  Finally, Section~\ref{sec:conclusions} summarizes our conclusions and discusses prospects for future multi-probe analyses.

\section{Cosmological Framework and Fiducial Parameters}
\label{sec:cosmo_theory}

In this section, we summarize the theoretical background relevant for interpreting DESI DR2 BAO measurements and clarify the role of the sound horizon calibration.

\subsection{Relation Between BAO Fitting Parameters and Distance Observables}
\label{subsec:bao_alpha_mapping}

The DESI BAO fitting pipeline constrains the anisotropic BAO signal through the scaling parameters
\begin{align}
\alpha_\parallel(z) &\equiv \frac{D_H(z)\, r_d^{\rm fid}}{D_H^{\rm fid}(z)\, r_d}, 
&
\alpha_\perp(z) &\equiv \frac{D_M(z)\, r_d^{\rm fid}}{D_M^{\rm fid}(z)\, r_d}, \label{alphapaper}
\end{align}
where $D_H(z)=c/H(z)$ is the radial Hubble distance, $D_M(z)$ is the comoving angular diameter distance, and $r_d$ is the sound horizon at the drag epoch. The superscript ``fid'' denotes the fiducial cosmology assumed in the pipeline. It is common to reparameterize these as
\begin{align}
\alpha_{\rm iso}(z) &\equiv \big[\alpha_\parallel(z)\,\alpha_\perp^2(z)\big]^{1/3}, 
&
\alpha_{\rm AP}(z) &\equiv \frac{\alpha_\parallel(z)}{\alpha_\perp(z)}, \label{alpha2}
\end{align}
where $\alpha_{\rm iso}$ captures the isotropic BAO scale and $\alpha_{\rm AP}$ the Alcock–Paczynski (AP) distortion ratio. DESI DR2 adopts the $(\alpha_{\rm iso},\alpha_{\rm AP})$ basis as baseline since they are nearly uncorrelated. Importantly, the BAO pipeline constrains ratios of distances to $r_d$ rather than absolute distances, so $r_d$ is not internally fixed by the BAO likelihood.

The $\alpha$-parameters map directly to distance observables as
\begin{align}
\frac{D_M(z)}{r_d} &= \alpha_\perp(z)\;\frac{D_M^{\rm fid}(z)}{r_d^{\rm fid}}, 
&
\frac{D_H(z)}{r_d} &= \alpha_\parallel(z)\;\frac{D_H^{\rm fid}(z)}{r_d^{\rm fid}}.
\label{eq:dm_dh_from_alpha}
\end{align}
Composite quantities follow immediately,
\begin{align}
\frac{D_V(z)}{r_d} 
&= \alpha_{\rm iso}(z)\;\left[\frac{D_M^{\rm fid\,2}(z)\;z\;D_H^{\rm fid}(z)}{r_d^{\rm fid\,3}}\right]^{1/3}, \label{eq:dv_from_alpha} \\[4pt]
\frac{D_M(z)}{D_H(z)} 
&= \frac{1}{\alpha_{\rm AP}(z)}\;\frac{D_M^{\rm fid}(z)}{D_H^{\rm fid}(z)}. \label{eq:dm_over_dh_from_alpha}
\end{align}
Thus the primary outputs of BAO fits are $\alpha_\parallel$ and $\alpha_\perp$, while $(D_V/r_d,\,D_M/D_H)$ are derived.

\subsection{Background Expansion and the CPL Framework}

DESI DR2 BAO observables are the transverse comoving distance $\DM(z)/r_d$, the radial Hubble distance $\DHh(z)/r_d$, the isotropic distance $\DV(z)/r_d$, and the dimensionless ratio $\DM(z)/\DHh(z)$ \cite{DESI:2025zgx}
\begin{align}
   & \DHh(z) = \frac{c}{H(z)} \,,\qquad
    \DM(z) = \frac{c}{H_0} \int_0^z \frac{dz'}{E(z')} = \frac{c}{100} \int_0^z \frac{dz'}{E h(z')} \,, \label{eq:DM_DESI} \\
   & \DV(z) = \left[ \DM(z)^2 \cdot \frac{c z}{H(z)} \right]^{1/3} \,,\qquad
    \frac{\DM(z)}{\DHh(z)} = E(z) \int_0^z \frac{dz'}{E(z')} = E h(z) \int_0^z \frac{dz'}{E h(z')} \,, \label{eq:DMoverDH_DESI}
\end{align}
where $H_0 = 100 h$ km/s/Mpc,  $E(z) = H(z)/H_0$, and $E h(z) = H(z)/100$.  

For the CPL parametrization~\cite{Chevallier:2000qy,Linder:2002et}, the dark-energy (DE) equation of state (e.o.s) and expansion history are
\begin{align}
&    \omega(z) = \oo + \wa \frac{z}{1+z} \,, \qquad
E(z) = \sqrt{
\Omegar (1+z)^4 + \Omegam (1+z)^3 + \Omegade (1+z)^{3(1+\oo+ \wa)} e^{-3\wa \frac{z}{1+z}}
} \,, \nonumber \\
& E h(z) = \sqrt{
\omega_{\rm r} (1+z)^4 + \omega_{\rm m} (1+z)^3 + \omega_{\rm de} (1+z)^{3(1+\oo+\wa)} e^{-3\wa \frac{z}{1+z}}
} \,, \label{EzEhz}
\end{align}
with spatial flatness imposing $\Omegade = 1 - \Omegam - \Omegar$ (equivalently, $\omega_{\rm de} = h^2 - \omega_{\rm m} - \omega_{\rm r}$). The radiation density today is given by
\begin{equation}
\Omegar = \Omega_{\gamma 0} \left[ 1 + \frac{7}{8} \left( \frac{4}{11} \, N_{\rm eff}  \right)^{4/3} \right] ,\quad
\Omega_{\gamma 0} h^2 \equiv \omega_{\gamma} \simeq 2.47 \times 10^{-5},  \label{Omegar0} %\ N_{\rm eff} = 3.046,
\end{equation}
where $\Omega_{\gamma 0}$ is the present photon density contrast and $N_{\rm eff}$ is the effective number of relativistic neutrino species.  

Thus, the DESI observables in the CPL framework are
\begin{align}
\frac{\DHh(z)}{r_d} &= \frac{c}{H(z) r_d} = \frac{c}{H_0 r_d} \frac{1}{E(z)}
= \frac{c}{100} \frac{1}{h r_d} \frac{1}{E(z)} = \frac{c}{100} \frac{1}{r_d} \frac{1}{Eh(z)} \label{extra1} \,, \\
\frac{\DM(z)}{r_d} &= \frac{c}{100} \frac{1}{h r_d} \int_0^z \frac{dz'}{E(z')} = \frac{c}{100} \frac{1}{r_d} \int_0^z \frac{dz'}{Eh(z')} \label{extra2} \,. 
\end{align}
Hence, BAO-only constraints effectively determine the combination $h\,r_d$ together with the shape of $E(z)$ set by $(\Omegam,\oo,\wa)$.

\subsection{Sound Horizon Calibration}

The comoving sound horizon at the drag epoch is often estimated using the DESI fitting formula~\cite{DESI:2025zgx,Brieden:2022heh},
\begin{equation}
r_d^{\rm DESI} = 147.05\,{\rm Mpc} \left(\frac{\omega_{\rm b}}{0.02236}\right)^{-0.13}
\left(\frac{\omega_{\rm m}}{0.1432}\right)^{-0.23}
\left(\frac{N_{\rm eff}}{3.04}\right)^{-0.1}\,,\label{eq:rd_fitting}
\end{equation}
where $\omega_{\rm b} = \Omega_{\rm b} h^2$ is related to the baryon energy density contrast.  This fitting form is widely used for internal consistency checks and parameter forecasts.  In our analysis, we consider two treatments: (i) BAO-only, in which $r_d$ is sampled self-consistently along with $(\omega_{\rm b},\omega_{\rm m},h)$, and (ii) BAO+$r_d$ prior, in which a Gaussian prior on $r_d$ from Planck DR3 is imposed. . The latter isolates the BAO-ruler calibration without using the full CMB likelihood and serves as our fixed-$r_d$ anchor for robustness tests, in contrast to combinations that effectively anchor $\theta_\ast$.  The CPL extensions considered in this work ($\wCDM$ and $\wowaCDM$) modify only the late-time expansion history through $E(z)$ and do not introduce changes to pre-recombination physics. In particular, the radiation density and $N_{\rm eff}$ are held fixed, and the sound horizon remains determined by the standard early-universe physics encoded in Eq.~(\ref{eq:rd_fitting}). Therefore, the Planck-derived Gaussian prior on $r_d$ is consistent within the scope of these models. The numerical value $r_d = 147.09 \pm 0.26$ Mpc corresponds to the Planck 2018 TT,TE,EE+lowE+lensing baseline result within flat $\Lambda$CDM~\cite{Planck:2018vyg}. We adopt this value as a precise early-universe calibration under standard recombination physics, consistent with the assumptions of the present analysis.

\subsection{Parameter Space}

Because the BAO likelihood constrains the ratios $\{D_H/r_d,\, D_M/r_d\}$, the natural BAO-only parameter vector is
\begin{equation}
\boldsymbol{\theta}_{\rm BAO}=\{\Omegam,\; h\,r_d,\; \oo,\; \wa\}. \label{eq:theta_bao}
\end{equation}
In BAO-only analyses, the observables $D_M/r_d$ and $D_H/r_d$ constrain only the product $h r_d$, leading to a pronounced degeneracy between $h$ and $r_d$ individually. As a result, the posterior distribution exhibits an elongated ridge in the $(h, r_d)$ plane. When a Gaussian prior on $r_d$ is imposed, this degeneracy is explicitly broken, allowing $h$ to be constrained independently. Figure~\ref{fig:hrd_degeneracy} illustrates this degeneracy structure and its collapse under the $r_d$ prior. Here $h\,r_d$ sets the absolute BAO ruler, while $(\Omegam,\oo,\wa)$ control the late-time expansion via $E(z)$ (flatness gives $\Omegade=1-\Omegam-\Omegar$ with $\Omegar$ fixed by $(\Omega_{\gamma0},N_{\rm eff})$). When we add a Gaussian prior on the sound horizon from Planck DR3, the $h$–$r_d$ degeneracy is broken by the external calibration. In this case it is more transparent to work with the physical matter density and the Hubble parameter,
\begin{equation}
\boldsymbol{\theta}_{\rm BAO+}r_d=\{\omega_m,\; h,\; \oo,\; \wa\} \,.  \label{eq:theta_bao_rd}
\end{equation}
Thus, BAO-only analyses constrain $\{\Omegam,\,h r_d,\,\oo,\,\wa\}$, whereas BAO combined with an $r_d$ prior effectively constrains $\{\omega_{\rm m},\,h,\,\oo,\,\wa\}$ (with $\Omegam$ obtained from $\omega_{\rm m}/h^2$). This separation cleanly distinguishes late-time dynamics from the externally imposed early-universe calibration.

\section{DESI DR2 BAO Data and Covariance Reconstruction}
\label{sec:data_theory}

To enable a direct model comparison of $\LCDM$, $\wCDM$, and $\wowaCDM$ using MCMC methods, we closely follow the procedure adopted in the DESI DR2 analysis.  In particular, we work with the anisotropic BAO observables $D_M/r_d$ and $D_H/r_d$,  which constitute the primary constraints in the DESI BAO likelihood pipeline and directly enter our parameter inference. This section describes the BAO observables and redshift binning, and details the reconstruction of the covariance matrix required for likelihood evaluation.

\subsection{DESI DR2 Observables and Redshift Bins}

DESI DR2 provides four types of BAO observables
\begin{equation}
 \mathcal{O}_i =  \left\{ D_V(z)/r_d,\; D_M(z)/r_d,\; D_H(z)/r_d,\; D_M(z)/D_H(z) \right\}, \label{Obs}
\end{equation}
reported across seven effective redshift bins, derived from a configuration-space analysis of galaxy clustering. 
Each observable is accompanied by its measurement uncertainty. 
A summary of the redshift bins and available observables is given in Table IV of Ref.~\cite{DESI:2025zgx}. 
For cosmological inference, we primarily employ the anisotropic pair $(D_M/r_d,\,D_H/r_d)$, consistent with the DESI likelihood construction and avoiding double counting with $D_V/r_d$. 

For clarity and completeness, we summarize here the subset of the DESI DR2 BAO dataset used in this work. We employ the anisotropic observables $(D_M/r_d, D_H/r_d)$ across seven effective redshift bins: $z = 0.295$ (BGS), $0.510$ (LRG1), $0.706$ (LRG2), $0.934$ (LRG3+ELG1), $1.321$ (ELG2), $1.484$ (QSO), and $2.330$ (Ly$\alpha$). The BGS bin provides only a $D_V/r_d$ measurement, while the remaining bins provide anisotropic constraints. To avoid double counting and maintain consistency with the DESI likelihood structure, we primarily use the $(D_M/r_d, D_H/r_d)$ basis, resulting in a total of 19 independent data points entering the likelihood.

\subsection{Covariance Matrix Reconstruction}

DESI DR2 provides observational uncertainties and partial correlation coefficients but does not release the full covariance matrix. 
We reconstruct a block-diagonal covariance matrix based on the reported uncertainties and correlation coefficients at each redshift, following the methodology of Ref.~\cite{Lee:2025kbn}. Each block corresponds to a redshift bin and contains up to three correlated observables. The resulting matrix covers 19 independent observables across six bins, excluding BGS, which contributes only a single $D_V/r_d$ measurement,  and is used for all likelihood-based calculations including $\chi^2$, AIC, and BIC statistics. While our reconstruction assumes Gaussian errors and neglects bin-to-bin correlations, it provides a consistent and reproducible basis for MCMC parameter estimation. 

To validate our covariance implementation, we use the publicly released uncertainties and correlation coefficients, restricting the analysis to the same BAO combinations considered in our likelihood construction. We verified that, within $\Lambda$CDM, the resulting BAO-only best-fit parameters are consistent with the official DESI DR2 results within the reported statistical uncertainties. This confirms that our covariance reconstruction faithfully reproduces the DESI likelihood structure and does not introduce significant systematic deviations in parameter inference.

Concretely, we organize the covariance into two independent $2\times 2$ blocks to avoid double counting, as in the DESI likelihood:
\begin{align}
C^{(z)} =
\begin{pmatrix}
\sigma^2_{V} & \rho_{V,M/H}\,\sigma_V\sigma_{M/H} \\
\rho_{V,M/H}\,\sigma_V\sigma_{M/H} & \sigma^2_{M/H}
\end{pmatrix}
\oplus
\begin{pmatrix}
\sigma^2_{M} & \rho_{M,H}\,\sigma_M\sigma_H \\
\rho_{M,H}\,\sigma_M\sigma_H & \sigma^2_H
\end{pmatrix}.
\label{CovM_block}
\end{align}
Here the first block corresponds to the $(D_V/r_d,\,D_M/D_H)$ basis, 
and the second to the $(D_M/r_d,\,D_H/r_d)$ basis. 
The explicit numerical covariance values for each redshift bin are listed below:
\begin{align}
&\textbf{Case 1: BGS (z=0.295) $D_V/r_d$} \nonumber \\
&C_{\text{BGS}} =
\begin{pmatrix}
0.005625
\end{pmatrix}  \nonumber
\\[1em]
&\textbf{Case 2: LRG1 (z=0.510)} \nonumber \\
&C^{(1)}_{\text{LRG1}} =
\begin{pmatrix}
9.801000\times 10^{-3} & 8.415000\times 10^{-5} \\
8.415000\times 10^{-5} & 2.890000\times 10^{-4}
\end{pmatrix},\quad
C^{(2)}_{\text{LRG1}} =
\begin{pmatrix}
2.788900\times 10^{-2} & -3.257752\times 10^{-2} \\
-3.257752\times 10^{-2} & 1.806250\times 10^{-1}
\end{pmatrix} \nonumber
\\[1em]
&\textbf{Case 3: LRG2 (z=0.706)} \nonumber \\
&C^{(1)}_{\text{LRG2}} =
\begin{pmatrix}
1.210000\times 10^{-2} & -4.158000\times 10^{-5} \\
-4.158000\times 10^{-5} & 4.410000\times 10^{-4}
\end{pmatrix},\quad
C^{(2)}_{\text{LRG2}} =
\begin{pmatrix}
3.132900\times 10^{-2} & -2.359764\times 10^{-2} \\
-2.359764\times 10^{-2} & 1.089000\times 10^{-1}
\end{pmatrix} \nonumber
\\[1em]
&\textbf{Case 4: LRG3+ELG1 (z=0.934)} \nonumber \\
&C^{(1)}_{\text{LRG3+ELG1}} =
\begin{pmatrix}
8.281000\times 10^{-3} & 9.682400\times 10^{-5} \\
9.682400\times 10^{-5} & 3.610000\times 10^{-4}
\end{pmatrix},\quad
C^{(2)}_{\text{LRG3+ELG1}} =
\begin{pmatrix}
2.310400\times 10^{-2} & -1.220377\times 10^{-2} \\
-1.220377\times 10^{-2} & 3.724900\times 10^{-2}
\end{pmatrix} \nonumber
\\[1em]
&\textbf{Case 5: ELG2 (z=1.321)} \nonumber \\
&C^{(1)}_{\text{ELG2}} =
\begin{pmatrix}
3.0276\times 10^{-2} & 1.58166\times 10^{-3} \\
1.58166\times 10^{-3} & 2.0250\times 10^{-3}
\end{pmatrix},\quad
C^{(2)}_{\text{ELG2}} =
\begin{pmatrix}
1.01124\times 10^{-1} & -3.050065\times 10^{-2} \\
-3.050065\times 10^{-2} & 4.8841\times 10^{-2}
\end{pmatrix} \nonumber
\\[1em]
&\textbf{Case 6: QSO (z=1.484)} \nonumber \\
&C^{(1)}_{\text{QSO}} =
\begin{pmatrix}
1.58404\times 10^{-1} & 2.38163\times 10^{-3} \\
2.38163\times 10^{-3} & 1.8496\times 10^{-2}
\end{pmatrix},\quad
C^{(2)}_{\text{QSO}} =
\begin{pmatrix}
5.7760\times 10^{-1} & -1.9608\times 10^{-1} \\
-1.9608\times 10^{-1} & 2.66256\times 10^{-1}
\end{pmatrix} \nonumber
\\[1em]
&\textbf{Case 7: Lya (z=2.330)} \nonumber \\
&C^{(1)}_{\text{Lya}} =
\begin{pmatrix}
6.5536\times 10^{-2} & 1.425357\times 10^{-2} \\
1.425357\times 10^{-2} & 9.4090\times 10^{-3}
\end{pmatrix},\quad
C^{(2)}_{\text{Lya}} =
\begin{pmatrix}
2.81961\times 10^{-1} & -2.311496\times 10^{-2} \\
-2.311496\times 10^{-2} & 1.0201\times 10^{-2}
\end{pmatrix} \nonumber
\end{align}

While our reconstruction assumes Gaussian errors and neglects bin-to-bin correlations, it provides a consistent and reproducible basis for MCMC parameter estimation.  This reconstructed covariance is employed consistently throughout our analysis: first for BAO-only model comparison, and subsequently when combined with a Planck-based Gaussian prior on $r_d$.  This ensures that differences between the two treatments originate solely from the choice of $r_d$ anchoring, not from covariance modeling.

\subsection{Validation of the Reconstructed Covariance}
\label{sec:cov_validation}

To validate the reconstructed covariance matrix, we perform a $\Lambda$CDM BAO-only MCMC analysis in the $(D_M/r_d, D_H/r_d)$ basis. The resulting posterior constraints are listed in Table~\ref{tab:pars-fit}, together with the official DESI DR2 $\Lambda$CDM values quoted by the collaboration~\cite{DESI:2025zgx}. For $\Omega_{m0}$ we obtain $0.2949^{+0.0096}_{-0.0093}$, compared to $0.2975\pm0.0086$ from DESI, and for $h r_d$ we obtain $101.80\pm0.84$ compared to $101.54\pm0.73$. The numerical differences are smaller than the quoted posterior uncertainties. This demonstrates that the reconstructed covariance reproduces the DESI DR2 BAO-only baseline at the level relevant for model comparison.

\section{Model Comparison: $\LCDM$, $\wCDM$, and $\wowaCDM$ with DESI DR2 BAO ($D_M/r_d$, $D_H/r_d$)}
\label{sec:cmp_lcdm_w_w0wa}

We compare three nested dark-energy models using DESI DR2 BAO observables in the $(D_M/r_d, D_H/r_d)$ basis,  with a two-element data vector per redshift bin. The sound horizon $r_d$ is absorbed into the free parameter $h\,r_d$ and is not independently constrained.  This setup cleanly isolates the constraining power of BAO distances alone, without anchoring to external CMB information. 

\subsection{Posterior constraints and fit quality}
\label{subsec:post-fit}

Table~\ref{tab:pars-fit} summarizes posterior constraints (medians with central 68\% credible intervals) and fit-quality statistics.
We also list the degrees of freedom (${\rm dof} = N_{\rm data}-N_{\rm params}$), reduced chi-square $\tilde\chi^2\equiv\chi^2/{\rm dof}$, and information criteria. The comparison with $\wCDM^{\textrm{DESI}}$ demonstrates that our independent MCMC analysis in the $(D_M/r_d, D_H/r_d)$ basis yields parameter constraints consistent with the DESI DR2 release within the quoted uncertainties.  Any small differences are well within the statistical precision of the data and do not affect the model-selection conclusions. The inclusion of the DESI-reported constraints allows a direct validation of our implementation and demonstrates consistency with the official DR2 analysis.

\begin{table*}[t]
\centering
\caption{Posterior constraints and fit quality for $\LCDM$, $\wCDM$, and $\wowaCDM$ using DESI DR2 BAO data in the $(D_M/r_d, D_H/r_d)$ basis, derived from our MCMC analysis. 
Reported values are medians with central 68\% credible intervals, while $\chi^2_{\min}$, reduced $\tilde{\chi}^2$, AIC, and BIC are given where available.  
Superscripts denote external references: $\LCDM^{\textrm{DESI}}$, $\wCDM^{\textrm{DESI}}$, and $\wowaCDM^{\textrm{DESI}}$ correspond to the parameters quoted directly by the DESI DR2 collaboration~\cite{DESI:2025zgx}; 
$\LCDM^{\textrm{Planck}}$ denotes Planck 2018 TT,TE,EE+lowE+lensing constraints~\cite{Planck:2018vyg}. 
For CPL, $\wowaCDM^{\textrm{I}}$ adopts a uniform prior $\wa \in [-2.5,\,2.5]$, while $\wowaCDM^{\textrm{II}}$ uses $\wa \in [-5.0,\,5.0]$.  
Note that in the DESI tables $\wa$ is reported as an upper limit, likely reflecting the HPD convention when the posterior is highly non-Gaussian.}
\label{tab:pars-fit}
\begin{ruledtabular}
\begin{tabular}{lcccccccc}
\toprule
Model & $\oo$& $\wa$ & $\Omegam$  & $h r_d$& $\chi^2_{\min}$ (dof) & $\tilde\chi^2$ & AIC & BIC  \\
\hline
\addlinespace[0.3em]
$\LCDM$ & -1 & 0 & $0.2949_{-0.0093}^{+0.0096}$&
$101.80_{-0.84}^{+0.84}$ &
$10.150\ (10)$ & $1.015$ & $14.149$ & $15.119$  \\
\addlinespace[0.3em]
$\LCDM^{\textrm{DESI}}$ & -1 & 0 & $0.2975\pm0.0086$&
$101.54\pm0.73$ & NA & NA & NA & NA  \\
\addlinespace[0.3em]
$\LCDM^{\textrm{Planck}}$ & -1 & 0 & $0.3142\pm0.0074$&
$99.24\pm0.82$ & NA & NA & NA & NA  \\
\addlinespace[0.5em]
$\wCDM$ & $-0.918^{+0.088}_{-0.093}$& 0 & $0.2964^{+0.0101}_{-0.0099}$ &
$99.87^{+2.32}_{-2.15}$ &
$9.470\ (9)$ & $1.052$ & $15.470$ & $16.925$  \\
\addlinespace[0.3em]
$\wCDM^{\textrm{DESI}}$ & $-0.916\pm0.078$& 0 & $0.2969\pm0.0089$ &
NA& NA & NA & NA & NA  \\
\addlinespace[0.5em]
$\wowaCDM^{\textrm{I}}$ & $-0.50^{+0.19}_{-0.32}$& $-1.56^{+1.13}_{-0.63}$& $0.352^{+0.024}_{-0.041}$ &$94.61^{+4.38}_{-2.80}$ &
$6.300\ (8)$ & $0.788$ & $14.300$ & $16.240$  \\
\addlinespace[0.3em]
$\wowaCDM^{\textrm{II}}$ & $-0.47^{+0.19}_{-0.33}$& $-1.66^{+1.16}_{-0.68}$& $0.356^{+0.024}_{-0.042}$ &$94.27^{+4.43}_{-2.54}$ &
$6.273\ (8)$ & $0.784$ & $14.273$ & $16.213$  \\
\addlinespace[0.3em]
$\wowaCDM^{\textrm{DESI}}$ & $-0.48^{+0.35}_{-0.17}$& $< -1.34$& $0.352^{+0.041}_{-0.018}$ &NA &
NA & NA& NA & NA  \\
\bottomrule
\end{tabular}
\end{ruledtabular}
\end{table*}

For CPL, the ($\oo$, $\wa$) posterior exhibits a pronounced degeneracy ridge.  Using equal-tailed percentiles we obtain $\wa = -1.56^{+1.13}_{-0.63}$ ($-1.66^{+1.16}_{-0.68}$) for different choice of uniform priors,  whereas adopting an HPD (highest posterior density) summary leads to a one-sided constraint, qualitatively consistent with the DESI-only bound (\textit{e.g.} $w_a < -1.34$ at 68\% credibility).  The physically constrained combination $\omega_p$ at $z_p \simeq 0.34$ remains close to $-1$, indicating that BAO data alone provide no statistically significant evidence for DDE. All three models provide statistically acceptable fits ($\tilde\chi^2\sim 0.8$--$1.05$).
The lower $\tilde\chi^2$ of $\wowaCDM$ is offset by its additional complexity, as reflected in the model-selection metrics below.

\subsection{Model selection: LRT, SDDR Bayes factor, and information criteria}
\label{subsec:model-sel}

For the nested comparisons $\LCDM$ $\subset$ $\wCDM$ (1 extra parameter) and $\LCDM$ $\subset$ $\wowaCDM$ (2 extra parameters), we evaluate the Wilks likelihood-ratio statistic (LRT), Savage--Dickey density ratio (SDDR), and $\Delta$AIC / $\Delta$BIC including model weights \cite{Jeffreys:1961,Liddle:2006kn,Liddle:2007fy,Trotta:2008qt}.

\begin{table*}[t]
\centering
\caption{Model-selection metrics relative to $\LCDM$.~$LRT$ reports $T_k=\chi^2_{\LCDM}-\chi^2_{\rm ext}$ with $k$ degrees of freedom; $p_{\rm LRT}$ is the corresponding tail probability. SDDR uses the posterior/prior density at the nested point ($w=-1$ for $\wCDM$; $(\oo, \wa)=(-1,0)$ for CPL). Akaike/BIC weights assume two-model comparison with equal priors.}
\label{tab:msel}
%\small
%\resizebox{0.7\textwidth}{!}{
\begin{ruledtabular}
\begin{tabular}{lccccccc}
\toprule
Extension & $T_k$ (dof) & $p_{\rm LRT}$ & $B_{01}$ & $\Delta$AIC & $\Delta$BIC & $W_{(AIC)}$ & $W_{(BIC)}$ \\
\hline
\addlinespace[0.3em]
%\midrule
$\wCDM$ & $0.679$ (1) & $0.4099$ & $4.555$ & $1.321$ & $1.806$ & 
$0.341$ & $0.288$ \\
\addlinespace[0.3em]
$\wowaCDM$& $3.849$ (2) & $0.1460$ & $2.543$ & $0.151$ & $1.121$ &
$0.481$ & $0.363$ \\
\bottomrule
\end{tabular} %}
\end{ruledtabular}
\end{table*}

We now examine each entry of the above table in more detail as follows.
\begin{itemize}
    \item $T_k$ (Likelihood Ratio Test statistic):
    Defined as $T_k = \chi^2_{\LCDM} - \chi^2_{\rm ext}$, with $k$ the number of additional parameters.  
    Under the null hypothesis it follows a $\chi^2_k$ distribution.  
    Rules of thumb:
    \begin{itemize}
      \item $T_k \lesssim 1$ (for $k=1$) or $\sim k$: negligible improvement, consistent with chance.
      \item $T_k \sim 2$--$4$ (for $k=1$): mild improvement, $p_{\rm LRT}\sim 0.05$--$0.2$, not significant.
      \item $T_k \gtrsim 6$ (for $k=1$) or $T_k \gtrsim 9$ (for $k=2$): corresponds to $p_{\rm LRT}<0.05$, usually considered significant.
    \end{itemize}
    $T_1=0.679$ ($\wCDM$) and $T_2=3.849$ ($\wowaCDM$) are both modest, showing no strong evidence for extra parameters.   
     \item $p_{\rm LRT}$ (tail probability):
    The probability that a $\chi^2_k$ random variable exceeds $T_k$.  
    Larger values mean the observed improvement can easily arise by chance.  
    \begin{itemize}
        \item $p_{\rm LRT}<0.05$: often called ``statistically significant,'' may support the extension. 
        \item $p_{\rm LRT}\gtrsim 0.1$: improvement likely due to chance; $\LCDM$ sufficient.
        \item $p_{\rm LRT}\sim 0.3$--$0.5$: essentially no improvement.
    \end{itemize}
    $p_{\rm LRT}=0.41$ ($\wCDM$) indicates pure chance improvement, while $p_{\rm LRT}=0.15$ ($\wowaCDM$) is somewhat stronger but still not significant.    
    \item $B_{01}$ (Bayes factor via SDDR (Savage-Dickey density ratio)):
    Ratio of posterior to prior density at the nested point ($\oo=-1$ for wCDM; $(\oo,\wa)=(-1,0)$ for $\wowaCDM$).  By convention, $B_{01}>1$ favors $\LCDM$.  
    Jeffreys’ scale:
    \begin{itemize}
      \item $1<B_{01}<3$: weak evidence for $\LCDM$,
      \item $3<B_{01}<10$: substantial evidence,
      \item $B_{01}>10$: strong to very strong evidence.
    \end{itemize}
$B_{01}=4.6$ ($\wCDM$) falls in the ``substantial'' range of the Jeffreys scale,  but does not constitute decisive evidence for $\LCDM$,  while $B_{01}=2.5$ ($\wowaCDM$) corresponds to weak evidence.
    
    \item \textbf{$\Delta$AIC, $\Delta$BIC:} 
    Information-criterion differences relative to $\LCDM$.  Values $\lesssim 2$ indicate comparable support.  
    Here $\Delta$AIC=$1.3$ ($\wCDM$) and $0.15$ ($\wowaCDM$); $\Delta$BIC=$1.8$ ($\wCDM$) and $1.1$ ($\wowaCDM$).  
Both criteria show only small differences relative to $\LCDM$ and do not provide decisive evidence favoring any model.
    
    \item \textbf{$W_{\rm (AIC)}$, $W_{\rm (BIC)}$ (Akaike/BIC weights):} 
    Defined as 
    \[
     W_i = \frac{\exp[-\tfrac{1}{2}\Delta_i]}{\sum_j \exp[-\tfrac{1}{2}\Delta_j]} ,
    \]
    where $\Delta_i = {\rm IC}_i - \min_j({\rm IC}_j)$.  
    These approximate posterior model probabilities assume equal priors and normalize to $\sum_i W_i=1$.  
    Values near $0.5$ imply no model is decisively favored.  
    In Table~II, $\wCDM$ has $(W_{\rm AIC},W_{\rm BIC})=(0.34,0.29)$ and $\wowaCDM$ $(0.48,0.36)$, showing that both remain viable, with $\LCDM$ retaining a slight edge.
\end{itemize}

Neither extension yields a statistically significant improvement over $\LCDM$. The LRT $p$-values ($0.41$ for $\wCDM$, $0.15$ for $\wowaCDM$) indicate that the observed improvements are compatible with statistical fluctuations. Bayes factors and information-criterion differences correspond to at most weak-to-moderate evidence on conventional scales, with all models remaining statistically viable.  LRT $p$-values ($0.41$ for $\wCDM$, $0.15$ for $\wowaCDM$) show no strong signal.  Bayes factors based on BIC approximations indicate at most weak differences between models, with no decisive support for any extension.

\subsection{Correlation structure, degeneracies, and pivot combination}
\label{subsec:rho-pivot}

The fitted parameters show structured (anti-)correlations shaped by Alcock--Paczyński geometry and the $\wowaCDM$ degeneracy ridge.

\paragraph{$\wCDM$ (3$\times$3).}
Using $(\Omega_{m0}, h r_d, w)$ we find
\begin{align}
\rho(\Omega_{m0},h r_d)=-0.471,\qquad \rho(h r_d,w)=-0.937,\qquad \rho(\Omega_{m0},w)=0.164. \label{wCDMzp}
\end{align}
The very strong anti-correlation between $h r_d$ and $\omega$ reflects the trade between the standard-ruler scale and late-time expansion history in matching $(D_M/r_d,D_H/r_d)$.

\paragraph{$\wowaCDM$ (4$\times$4).}
Strong CPL anti-correlation $\rho(\oo,\wa)=-0.935$ defines an elongated likelihood ridge with major-axis angle $\simeq104.4^\circ$ in the $(\oo,\wa)$ plane.  The pivot redshift is defined as the redshift at which the uncertainty on the dark energy equation of state $w(a)$ is minimized. For the CPL parameterization,
\begin{equation}
w(a) = w_0 + w_a (1-a),
\end{equation}
the variance of $w(a)$ is
\begin{equation}
\mathrm{Var}[w(a)] = 
\mathrm{Var}(w_0)
+ (1-a)^2 \mathrm{Var}(w_a)
+ 2(1-a)\,\mathrm{Cov}(w_0,w_a).
\end{equation}
Minimizing this expression with respect to $a$ yields the pivot scale factor
\begin{equation}
a_p = 1 + \frac{\mathrm{Cov}(w_0,w_a)}{\mathrm{Var}(w_a)}.
\end{equation}
The corresponding pivot redshift is then
\begin{equation}
z_p = \frac{1}{a_p} - 1.
\end{equation}
The value $z_p \simeq 0.34$ quoted in the text is obtained directly from the covariance matrix of $(w_0, w_a)$ derived from the posterior MCMC chains. Decorrelating via the standard pivot 
\begin{align}
\omega_p = \oo + (1-a_p) \wa,\qquad a_p=0.7455\ (z_p\simeq0.341), \label{omegap}
\end{align}
yields a directly constrained combination 
\begin{align}
\omega_p = -0.899 \pm 0.087,
\end{align}
consistent with $-1$ at only $\simeq 1.2\sigma$, reinforcing that the pivoted constraint is fully compatible with $\LCDM$.

\subsection{Posterior predictive checks (PPC)}
\label{subsec:ppc}

We compute discrepancy-based PPC $p$-values using the $\chi^2$ statistic:
\begin{align}
p_{\rm PPC}\equiv P(\chi^2_{\rm rep}\ge\chi^2_{\rm obs}\mid{\rm model}). \label{pPPC}
\end{align}
For $\wCDM$, $p_{\rm PPC}=0.662$; for $\wowaCDM$, $p_{\rm PPC}=0.891$. Both are comfortably away from the extremes, indicating that replicated data typically produce comparable or larger discrepancies than observed—hence there is no evidence of overfitting and the $\LCDM$ baseline remains statistically sufficient.

While posterior predictive $p$-values substantially larger than $0.5$ indicate that the observed data are well reproduced by the model, values approaching unity can sometimes raise concerns about excessive model flexibility or insufficient sensitivity of the discrepancy statistic. Therefore, we examine whether the relatively high value obtained for $\omega_0\omega_a$CDM reflects overfitting. First, the PPC statistic employed here is identical across all models, ensuring a fair comparison. Second, model complexity is independently accounted for through information criteria (AIC and BIC), which penalize additional parameters. The corresponding AIC/BIC values do not indicate an excessive preference for the more flexible model. Taken together, the PPC results suggest that the observed discrepancies are typical realizations under the model rather than evidence of overfitting.

\subsection{Synthesis}
\label{subsec:synthesis}

Across complementary diagnostics (LRT, SDDR, AIC/BIC, PPC), the present BAO-only $(D_M/r_d,D_H/r_d)$ analysis exhibits at most a weak preference for $\LCDM$ over $\wCDM$ and $\wowaCDM$.  $\wowaCDM$'s lower $\chi^2$ is balanced by complexity penalties, while its pivoted constraint $\omega_p$ remains statistically consistent with $-1$.  The dominant limitation is discussed further in Sec.~\ref{sec:results}, where we highlight the role of degeneracies between $h r_d$ and dark-energy parameters.

%------------------------------------------------------------
\subsection{Degeneracy Between $h$ and $r_d$}
%------------------------------------------------------------

BAO observables entering our analysis are expressed in the form $D_M(z)/r_d$ and $D_H(z)/r_d$.  Neglecting radiation at BAO redshifts, these quantities scale approximately as
\begin{equation}
\frac{D_M(z)}{r_d}, \ \frac{D_H(z)}{r_d} \propto \frac{1}{h\, r_d},
\label{eq:hrd_scaling}
\end{equation}
up to a mild dependence on $\Omega_{m0}$ through the expansion rate $E(z)$. As a consequence, BAO-only likelihoods primarily constrain the product $h r_d$, rather than the individual parameters $h$ and $r_d$. This induces a strong degeneracy direction in the $(h, r_d)$ plane, approximately following contours of constant $h r_d$. When $r_d$ is treated as a free parameter, the posterior distribution therefore exhibits an extended ridge-like structure.

\begin{figure}[t]
\centering
\includegraphics[width=0.65\textwidth]{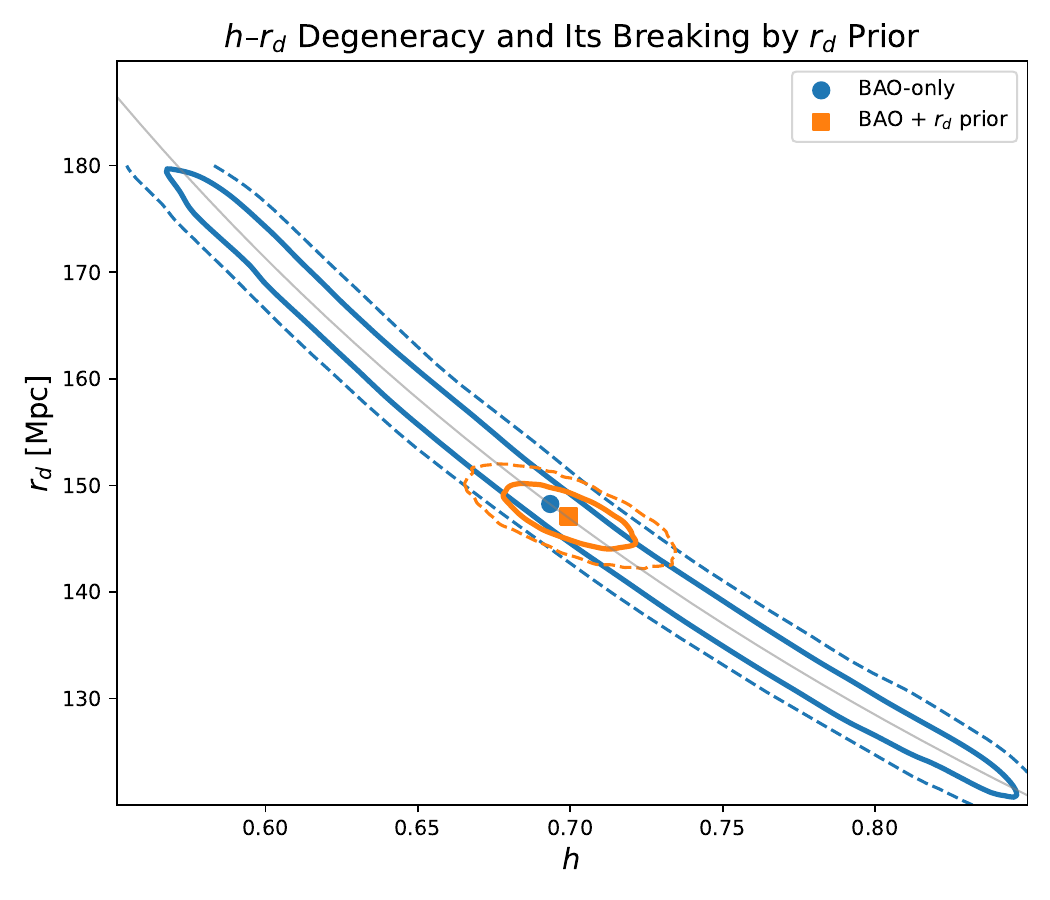}
\caption{Posterior contours in the $(h, r_d)$ plane. Solid lines denote the 68\% confidence regions (1$\sigma$), and dashed lines denote the 95\% confidence regions (2$\sigma$). The BAO-only case exhibits a long ridge-like degeneracy approximately aligned with constant $h r_d$. When a Gaussian prior on $r_d$ is applied, this degeneracy is broken and the allowed region collapses into a compact elliptical shape.}
\label{fig:hrd_degeneracy}
\end{figure}

Figure~\ref{fig:hrd_degeneracy} illustrates this behavior. In the BAO-only case, the posterior contours form a long, elongated ridge aligned approximately with constant $h r_d$. When a Gaussian prior on $r_d$ is introduced, this degeneracy is broken and the posterior becomes localized, forming a compact elliptical region. In this case, the constraint on $h$ tightens correspondingly. This structure motivates the parameter choice adopted in Eq.~(13), where $h r_d$ is used as a fundamental BAO-only parameter. Once an external prior on $r_d$ is included, it becomes natural to treat $h$ and $r_d$ separately, as in Eq.~(14), since the degeneracy is then resolved.

\section{Model Comparison with Fixed $r_d$}
\label{sec:results}

We now assess the impact of including external early-universe information by imposing a Gaussian prior on the sound horizon from Planck DR3, $r_d = 147.09 \pm 0.26$ Mpc.  This strategy provides a controlled way to break the $(h r_d, \oo, \wa)$ degeneracy while avoiding the full complexity of the CMB likelihood. This approach differs from the DESI methodology, which incorporates the full CMB likelihood by treating $(\theta_\ast, \omega_b, \omega_m)$ as purely early-time quantities independent of $(\oo, \wa)$.  
In practice, shifts in $\Omegam$ and $h$ can alter both $r_d$ and $r_\ast$, thereby modifying the angular acoustic scale $\theta_\ast$ and indirectly coupling late-time parameters to early-universe observables.  
Table~\ref{tab:bestfit_rd_rstar} illustrates this point: the $\wowaCDM$ model yields a significantly higher $\Omegam$, which modifies $r_d$ and $r_\ast$ relative to $\LCDM$, even though $\oo$ and $\wa$ do not explicitly enter recombination physics.  This demonstrates that the common assumption of strict independence is only approximate; our use of an $r_d$ prior therefore isolates the impact of sound-horizon calibration in a transparent and reproducible manner.

\begin{table}[htbp]
\centering
\caption{Best-fit cosmological parameters and sound horizons obtained with CLASS~\cite{Blas:2011rf} at the DESI DR2 BAO best-fit values.}
\begin{ruledtabular}
\begin{tabular}{lccc}
\toprule
Model & $(\oo, \wa)$ & $(\Omegam, H_0$ [km/s/Mpc]) & $(r_d, r_\ast)$ [Mpc] \\
\hline
$\LCDM$   & $(-1.0, 0)$       & $(0.3027,\; 68.17)$ & $(147.22,\; 144.77)$ \\
$\wCDM$         & $(-1.055, 0)$     & $(0.2927,\; 69.51)$ & $(146.27,\; 144.10)$ \\
$\wowaCDM$    & $(-0.43, -1.70)$  & $(0.352,\; 63.7)$   & $(149.23,\; 145.78)$ \\
\bottomrule
\end{tabular}
\end{ruledtabular}
\label{tab:bestfit_rd_rstar}
\end{table}

\subsection{Posterior constraints and fit quality}

Table~\ref{tab:model_comparison_rd} summarizes the MCMC results using DESI DR2 BAO observables combined with the \textit{Planck}-based $r_d$ prior.  
All three models provide good fits, with reduced chi-square values $\tilde{\chi}^2 \simeq 0.9$--$1.2$.  
The $\wowaCDM$ model achieves the lowest $\chi^2$, but the difference relative to $\LCDM$ is small compared with the penalty for additional parameters.  As in the BAO-only case, the CPL parameters follow a strong degeneracy ridge, and the pivoted e.o.s remains consistent with $\omega=-1$ within $1.2\sigma$, confirming the absence of significant evidence for DDE.

\begin{table}[htbp]
\centering
\caption{Posterior constraints and fit quality for $\LCDM$, $\wCDM$, and $\wowaCDM$ using DESI DR2 BAO data combined with the Planck DR3 $r_d$ prior. For $\wowaCDM^{\textrm{I} (\textrm{II})}$ we adopt a relaxed prior $\wa\in [-2.5, +2.5] ([-5,+5])$.  The Gaussian prior on the sound horizon, $r_d = 147.09 \pm 0.26$ Mpc, is identical for all models. Any small differences in the reported posterior intervals (e.g., $147.09^{+0.25}_{-0.26}$) arise solely from MCMC sampling and rounding effects after combining the prior with the BAO likelihood, and do not reflect different prior assumptions.}
\begin{ruledtabular}
\begin{tabular}{lcccccccc}
\toprule
Model & $\oo$ & $\wa$ & $\Omegam$ & $h$ & $r_d$ [Mpc] & $\chi^2_{\min}$ (dof) & $\tilde\chi^2$ & AIC / BIC \\
\hline
$\LCDM$   & $-1$ & $0$ & $0.295\pm0.010$ & $0.692\pm0.006$ & $147.09\pm0.26$ & 10.16 (9) & 1.13 & 16.16 / 17.61 \\
$\wCDM$         & $-0.92\pm0.09$ & $0$ & $0.296\pm0.010$ & $0.679\pm0.016$ & $147.09\pm0.26$ & 9.48 (8) & 1.19 & 17.48 / 19.42 \\
$\wowaCDM^{\textrm{I}}$    & $-0.50^{+0.19}_{-0.32}$ & $-1.57^{+1.14}_{-0.62}$ & $0.352^{+0.024}_{-0.041}$ & $0.643^{+0.030}_{-0.019}$ & $147.09\pm0.26$ & 6.29 (7) & 0.90 & 16.29 / 18.71 \\
$\wowaCDM^{\textrm{II}}$    & $-0.47^{+0.19}_{-0.33}$ & $-1.67^{+1.15}_{-0.68}$ & $0.356^{+0.024}_{-0.041}$ & $0.641^{+0.030}_{-0.017}$ & $147.09^{+0.25}_{-0.26}$ & 6.25 (7) & 0.89 & 16.25 / 18.68 \\
\bottomrule
\end{tabular}
\end{ruledtabular}
\label{tab:model_comparison_rd}
\end{table}

\subsection{Model selection}

Table~\ref{tab:model_comparison2_rd} reports the model-selection metrics relative to $\LCDM$.  For $\wCDM$ the likelihood-ratio statistic is $T_1\simeq0.68$ with $p\simeq0.41$, while for $\wowaCDM$ it is $T_2\simeq3.87$ with $p\simeq0.14$.  
Neither reaches conventional significance.  Bayes factors based on BIC approximations are below unity, indicating weak support for $\LCDM$.  AIC values show a slight preference for $\wowaCDM$, but BIC imposes stronger complexity penalties, leaving $\LCDM$ marginally favored overall and the extended models statistically viable but not compelling.

\begin{table}[htbp]
\centering
\caption{Model-selection metrics relative to $\LCDM$ for DESI DR2 + Planck $r_d$ prior. For $\wowaCDM^{\textrm{I}}$ and $\wowaCDM^{\textrm{II}}$ we adopt different uniform priors on $\wa$,  namely $[-2.5, +2.5]$ and $[-5, +5]$, respectively.}
\begin{ruledtabular}
\begin{tabular}{lccccccc}
\toprule
Extension & $T_k$ (dof) & $p_{\rm LRT}$ & $B_{01}$ & $\Delta$AIC & $\Delta$BIC & $W_{\rm AIC}$ & $W_{\rm BIC}$ \\
\hline
\addlinespace[0.3em]
$\wCDM$       & 0.68 (1) & 0.41 & 0.41 & 1.32 & 1.81 & 0.21 & 0.20 \\
$\wowaCDM^{(\textrm{I})}$  & 3.87 (2) & 0.14 & 0.58 & 0.13 & 1.10 & 0.38 & 0.29 \\
\hline
\addlinespace[0.3em]
$\wCDM$       & 0.678 (1) & 0.410 & 2.467 & 1.322 & 1.806 & 0.341 & 0.288 \\
$\wowaCDM^{(\textrm{II})}$  & 3.907 (2) & 0.142 & 1.701 & 0.093 & 1.063 & 0.488 & 0.370 \\
\bottomrule
\end{tabular}
\end{ruledtabular}
\label{tab:model_comparison2_rd}
\end{table}

\subsection{Synthesis}

When combined with the \textit{Planck} $r_d$ prior, DESI DR2 BAO data remain consistent with $\LCDM$.  
The extended models provide slightly lower $\chi^2$ but do not achieve statistically significant improvements, and all model-selection metrics point to at most weak differences.  Importantly, the apparent $\sim3\sigma$ preference for DDE reported by the DESI collaboration (based on the full CMB likelihood) is not reproduced here. Instead, the data show that once the sound horizon calibration is imposed without double counting late-time information, no strong evidence for DDE emerges.  This highlights the sensitivity of DE inference to how early-universe information is incorporated, and underscores the value of fixed-$r_d$ analyses as a robustness test.  Similar sensitivity to anchor choice has also been discussed in eBOSS Ly$\alpha$ BAO studies~\cite{eBOSS:2020tmo},  and in SNe systematics analyses~\cite{Pan-STARRS1:2017jku}.

We note that in the BAO+$r_d$ prior analysis, the $\wowaCDM$ model yields a lower inferred value of $H_0$ relative to $\Lambda$CDM. This shift originates from internal parameter degeneracies between $(\Omegam, h, \oo, \wa)$ required to maintain consistency with the BAO distance ratios under fixed sound-horizon calibration. However, our analysis does not include late-time distance-ladder measurements such as SH0ES, nor does it incorporate the full Planck CMB likelihood. Therefore, this shift in $H_0$ should not be interpreted as a resolution of the Hubble tension. Rather, it reflects the geometric rearrangement of parameter space within the BAO likelihood under extended dark-energy models.

\section{Conclusions}
\label{sec:conclusions}

We have reassessed the cosmological implications of the DESI DR2 BAO measurements using the anisotropic
observables $(D_M/r_d, D_H/r_d)$, the primary outputs of the BAO fitting pipeline.  
By mapping the fitted $\alpha$-parameters to distance ratios and reconstructing a block-diagonal covariance per redshift bin, 
we carried out a uniform MCMC analysis of three nested models, $\LCDM$, $\wCDM$, and $\wowaCDM$.

In the BAO-only analysis, all three models provide statistically acceptable fits with reduced chi-square values 
$\tilde\chi^2\simeq0.8$--$1.05$. Neither $\wCDM$ nor $\wowaCDM$ yields a significant improvement over $\LCDM$: 
likelihood-ratio tests give $p_{\rm LRT}=0.41$ and $0.15$ respectively, while Bayes factors and information criteria mildly favor the simpler $\LCDM$ model.  The pivoted e.o.s, $\omega_p=-0.899\pm0.087$ at $z_p\simeq0.34$, is fully consistent with $-1$, and posterior predictive checks confirm the absence of overfitting.  These conclusions are in full agreement with the official DESI DR2 analysis.

When a Gaussian prior on $r_d = 147.09 \pm 0.26$ Mpc from Planck DR3 is imposed, the results remain consistent with the BAO-only case. Although $\wowaCDM$ achieves a somewhat lower $\chi^2$, the gain is modest and outweighed by complexity penalties, with model-selection metrics continuing to prefer $\LCDM$.  Importantly, the apparent $\sim3\sigma$ preference for DDE reported by DESI when using the full CMB likelihood is not reproduced under this more conservative treatment.  This demonstrates that conclusions about dark-energy dynamics depend sensitively on how CMB information is incorporated,  and that isolating the $r_d$ calibration provides a more transparent robustness test~\cite{Birrer:2020tax,Dutta:2018vmq}.

A common feature of both analyses is the strong degeneracy between $h\,r_d$ and dark-energy parameters, especially the anti-correlation with $\omega$.  BAO data alone therefore constrain $h\,r_d$ and geometric combinations of $E(z)$, but cannot disentangle late-time dynamics from early-time calibration without external information.

In summary, DESI DR2 BAO data, whether analyzed alone or combined with the \textit{Planck} $r_d$ prior, remain fully consistent with $\LCDM$ and show no statistically significant preference for $\wCDM$ or $\wowaCDM$. Importantly, the apparent $\sim3\sigma$ preference for dynamical dark energy reported by DESI when using the full CMB likelihood is not reproduced under our fixed-$r_d$ anchoring.

Our results should therefore be interpreted as a robustness test of the sound-horizon calibration rather than as a refutation or replacement of full CMB-based joint analyses. Future progress will hinge on achieving a precise and independent calibration of $r_d$, which is essential for breaking the degeneracy between $h\,r_d$ and dark-energy parameters.

Although the present work focuses exclusively on DESI DR2 BAO and BAO+$r_d$ prior combinations, it is useful to place these results in a broader observational context. Joint analyses combining BAO with supernova datasets (e.g., Pantheon+) have reported varying levels of preference for dynamical dark energy, depending on dataset selection and early-universe anchoring choices. Our BAO-based results therefore provide a complementary perspective by isolating the geometric information contained in BAO distance ratios from additional late-time probes. A full multi-probe analysis including SNe Ia or GRBs is beyond the scope of this work.

Several limitations of the present analysis should be emphasized. First, our results are restricted to BAO-only and BAO+$r_d$ prior configurations and do not incorporate the full Planck CMB likelihood. Second, the sound horizon is treated either as a fixed calibration parameter or as an external Gaussian prior; we do not explore early-universe modifications that alter recombination or the acoustic scale directly. Third, our conclusions apply specifically to the robustness of late-time dynamical dark-energy inference within the DESI BAO framework and should not be interpreted as a general statement about all joint CMB+LSS analyses.

%\acknowledgments
%SL is supported by the National Research Foundation of Korea (NRF) grant funded by the Ministry of Science and ICT (Grant No. RS-2021-NR059413).

\end{document}